\documentclass{ws-ijmpb}
\usepackage{graphicx}
\usepackage{epsf}
\begin{document}

\markboth{Ratnadeep Roy and Purusattam Ray}
{Dynamics of Random Field Ising \ldots}

%
\catchline{}{}{}{}{}
%

\title{RESPONSE OF RANDOM FIELD ISING MODEL DRIVEN BY AN  EXTERNAL FIELD} 

\author{RATNADEEP ROY$^\dagger$ and PURUSATTAM RAY$^\ddagger$} 
\address{The Institute of Mathematical Sciences,\\Chennai-600 113 ,\\India.
\\$^\dagger$deep@imsc.res.in\\
$^\ddagger$ray@imsc.res.in}

\maketitle

\begin{history}
\received{23 June 2003}
\end{history}

\begin{abstract}
We study the dynamics of spin flipping at first order transitions 
in zero temperature two-dimensional random-field Ising model 
driven by an external field. We find a critical value of the disorder 
strength at which a discontinuous sharp jump in magnetization
first occurs. We discuss growth morphology of the flipped-spin domains at 
and away from criticality.
\end{abstract}


\section{Introduction}  
The {\it Random Field Ising Model} $(RFIM)$ has been widely studied  
in recent years in the context of hysteresis [1-3], return-point memory [4,5]
and Barkhausen noise [6-11].  
Hysteresis is a common phenomenon, exhibited, for example, by most 
magnetic materials. In hysteresis the magnetization lags behind the applied 
field and the phenomenon has received lots of attention in the past. Moreover, 
if one observes carefully, the change in magnetization in systems like
 that in ferromagnetic alloys and amorphous ferromagnets, takes place in a 
series of irregular pulses as the external field is slowly varied. 
In experiments these pulses manifest themselves as acoustic emissions and is 
known as Barkhausen noise.  
Interestingly, Barkhausen signals show scale-free behavior similar to 
that of systems at criticality. 
Barkhausen noise has been studied much in recent years in the context
of driven disordered systems far from equilibrium example of which 
include martensite transformation in shape-memory alloys [12], eartquakes 
[13], deformation of granular materials [14] and the breakdown of 
solids [15]. 
Various experiments [8] show that the 
distribution of size, duration and energy associated with Barkhausen noise 
show power-law behavior. Sethna et al, in order to account for the 
hysteresis and Barkhausen noise incorporated disorder in the 
form of random fields into the otherwise pure spin systems such as the  
Ising model [3].    

\par In the {\it RFIM}, the Ising spins $\{s_i\}$ with nearest neighbor 
ferromagnetic interaction {\it J} are coupled to the on-site quenched 
random magnetic fields $\{h_i\}$ and the external field $h_a$. The 
Hamiltonian of the system is 

\begin{equation}
H = -J \sum_{<ij>}{s_i}{s_j} - \sum_{i}{h_i}{s_i} - {h_a}\sum_{i}{s_i}
\end{equation}

\noindent where $\{h_i\}$ are quenched independent random 
variable drawn from a distribution of zero mean and variance $\Delta$.
In the study of hysteresis and Barkhausen jumps the 
applied field is ramped up or down adiabatically (so that the rate 
of spin flips is much larger than the rate of change of $h_a$).     
In the absence of disorder ($\{h_i\} = 0$), the model exhibits a 
sharp first order transition taking the magnetization $m(h_a)$
from  -1 to +1 as $\it h_a$ passes 
through {\it zJ}, {\it z} being the coordination number of the underlying 
lattice. In presence of random field, the sharpness of the transition 
gets smeared out and the change in magnetization with field takes 
place in a series of ``spin-cluster'' flips and the magnetization 
changes in sporadic jumps as is observed in {\it BN}.

\par Simulations of {\it RFIM} in three-dimensions show that for   
large values of $\Delta$  
(greater than a critical value $\Delta_c$), the spins flip in 
small clusters resulting in avalanches of small sizes. 
The sizes of these avalanches are distributed in a power-law. At 
$\Delta = \Delta_c$ avalanches of all sizes occur. For $\Delta < \Delta_c$, 
the magnetization shows a first order jump (corresponding to an infinite 
avalanche) as is the case if there is no disorder. 

The behavior of avalanches and avalanche size distribution  
at and near $\Delta_c$ has been studied in the mean 
field limit [3]
and the power law relating to the avalanche size distribution is 
known exactly in {\it RFIM} in one-dimension and on Bethe lattices [11].
For a linear chain and a Bethe lattice of coordination number z =3 the 
avalanche size (s) disribution decreases exponentially for large {\it s}. 
No first order jump has been observed for any finite amount of disorder 
(unbounded) for $z \leq 3$. Thus the low $\Delta$ behavior depends 
on the coordination number of the underlying lattice. 
For $z \geq 4$ and for small disorder 
the magnetization shows a first order discontinuity for several continuous and 
unimodal distributions of the random fields. The avalanche distribution 
{\it P(s)} varies as $s^{-\frac{5}{2}}$ for large s near the 
discontinuity.  


\par This growth of 
magnetization is essentially due to the motion of an interface of a 
cluster of 
positive spins through a disordered media.  
Renormalization group  studies of the interface (starting from a continuum 
model and an interface without any backbends) motion in disordered 
medium shows that the interface at the critical value of the strength of 
the disorder becomes rough with a roughness exponent
$\zeta = (5-d)/3$ [16] for a d-dimensional system. According to this 
relation 
the upper critical dimension 
is five and the lower critical dimension is two. At $d = 2$, it is considered 
that the interface may have backbends and can encompass bubbles as it moves 
through the medium.
Simulation results show the following generic features of the interface 
growth in random media [17-24]: (i)  For large strengths of the disorder 
the 
interface growth is percolation-like.The fractal dimension of the 
magnetic-interface corresponds to that ordinary site 
percolation. (ii) At the critical value of the disorder the interface becomes 
self-affine with a roughness exponent which closely matches the RG result.
(iii) For low strength of the disorder the interface growth is faceted.  

\par We study  the spin flipping dynamics in {\it RFIM} in two-dimensions 
at zero temperature as the external field is ramped up 
slowly from a high negative value. Our intention is to see if there is any 
finite value of the disorder strength $\Delta_c$ in two-dimension. 
At $\Delta_c$ and at a particular field $h_a = h_{a}^{*}$, the spin starts 
flipping from a site and does not end till the flipped spins span the entire 
system.
The domain wall that separates the flipped spin regions from the rest is 
highly tortuous (fractal like). The spin flipped region has bubbles of 
unflipped spins, so that $m(h_a)$ does not jump from -1 to +1,
 instead attains a value $\sim 0.76$ irrespective of system size. In the regime
 when the flipped spin domain grows, the number {\it M} of flipped spins and 
the interface length {\it I}
both grow with time {\it t}  as $M \sim I \sim  t^2$. For $\Delta < \Delta_c$, 
{\it M}
jumps from -1 to +1 at a particular field and the flipped spin domain remains 
compact as it grows following $M \sim I^2 \sim t^2$. 
For $\Delta > \Delta_c$, $m(h_a)$ grows continuously with the field in 
jumps of 
different sizes. These jumps correspond to flipping of small spin clusters at 
various places in the lattice.

\section{The Model and Simulation}
We have studied RFIM in  two-dimensions on a square lattice of size 
{\it L} with 
a classical Ising spin $s_i = \pm 1$ and a quenched random field 
$h_i$ on every site $i$ of the lattice. Each spin interacts with 
each of its nearest neighbor spin through a ferromagnetic interaction 
$J = 1$. The random field $h_i$ is drawn from a uniform distribution

\begin{eqnarray}
p(h_i)  &=& {} \frac{1}{2\Delta}\qquad {\textrm{if}  -\Delta \leq h_i 
\leq\Delta}\nonumber \\
        &=& ~  0 \qquad~~~  {\textrm{otherwise .}}
\end{eqnarray}

\par The spins are subjected to an applied field $h_a$ which is varied. We 
study the model at zero temperature. We start with a configuration of 
all down spins which corresponds to a high negative value of $h_a$.
The field $h_a$ is then slowly raised and the resulting spin flips are 
recorded. The spin flipping process follows the rule of zero temperature 
Glauber dynamics [25], where a spin is flipped only if the flipping lowers 
the energy of the system as calculated according to the Hamiltonian given 
by ``Eq (1)".
\par We use periodic boundary conditions in our simulation. For a certain 
{\it L}, $\Delta$ and a configuration of the random 
field $h_i$ we first find out the spins $\{\alpha\}$ which can be most easily
flipped and the corresponding value of the applied 
field $h_{a}^{*}$ required to flip the spins. We then set $h_a = h_{a}^{*}$ 
and flip all the spins in 
$\{\alpha\}$ simultaneously. This  constitutes one time step in 
our simulation. 
Next we check if this primary set of spin 
flips introduce secondary spin flips and so on, always flipping all the 
flippable spins simultaneously. If there is no spin that can be flipped, 
we increase $h_a$ to trigger next generation of spin flipping. We continue 
this process till all the spins of the lattice are flipped. We have taken 
results for {\it L} = 500 to 7000 and for various values $\Delta$ from 
2.0 to 2.8. For 
a certain $\Delta$, our results are averaged over 50 initial random 
configurations of the on-site fields.

\section{Results and Discussion}
\par Our simulation results show that there is a critical value 
$\Delta_c$ of the strength of the disorder. The growth morphology of domains
of flipped spins are distinctly different for $\Delta < \Delta_c$,
$\Delta > \Delta_c$ and for $\Delta = \Delta_c$. We discuss the three
cases below:
\begin{romanlist}
\item For  $\Delta < \Delta_c$ as the external field is ramped up there 
is a sharp
first order transition (like that happens in the absence of disorder) at a 
particular 
field ($h_a = 4J - \Delta$)that takes the magnetization from -1 to +1 
as is shown in Fig.1
This infinite avalanche nucleates from a single spin and invades the system 
in course of time. The flipped spin cluster remains compact while it grows:
the mass {\it M} (the number of upturned spins) of the cluster
grows with time {\it t} (in Monte Carlo steps) as $ M \sim t^2$ (Fig.2). 
The number of spins
{\it I} along the  interface grows as $I \sim t$ (Fig.2).

\begin{figure}[th]
\vspace{1cm}
\centerline{\psfig{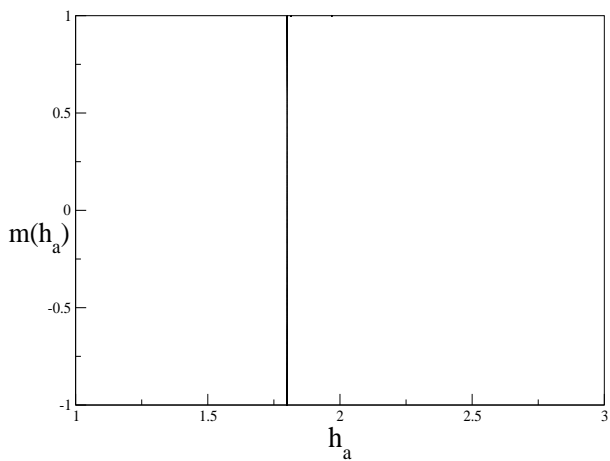}}
\vspace*{8pt}
\caption {The magnetization curve for $\Delta (=2.2) < \Delta_c $.
The magnetization shows a discontinuous jump at 
the critical field $h_a = 4J - \Delta$.}
\end{figure}

\begin{figure}[th]
\vspace{1cm}
\begin{center}
\includegraphics[width=8cm,height = 6cm]{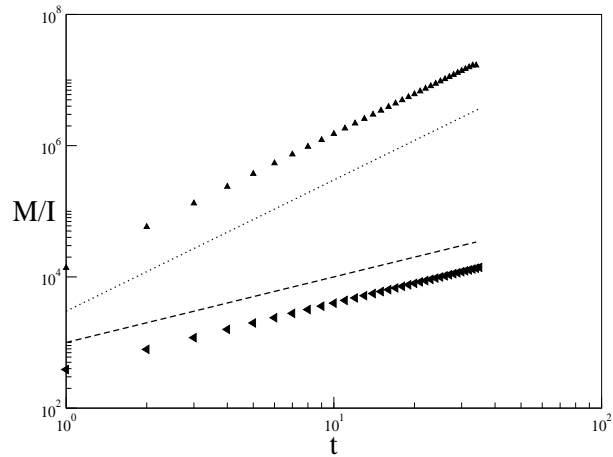}
\vspace{8pt}
\caption{The variation of the total number of spins,{\it M} and the number 
of spins along the interface I with time {\it t} (in units of
300 Monte Carlo steps) for $\Delta < \Delta_c = 
2.2$ and $h_a = h_{a}^{*} = 1.800$. {\it M } varies as the square of the 
time and {\it I} varies linearly with time. The dotted line has a slope 
of two and the dashed line has a slope of unity  and both
are a guide to the eye.}
\end{center}
\end{figure}

\item For  $\Delta > \Delta_c$ an infinite avalanche never happens. The 
magnetization increases in irregular steps  with  the 
increase of external field as shown in Fig.3. In Fig.4 we show a snapshot of 
the upturned spins at a particular field when the flipping of the spins
has stopped. In this regime we see nucleation of many domains as opposed to 
the nucleation of a single domain for $\Delta < \Delta_c$. Such a feature
is also observed in dilute {\it RFIM} [10]. 

\begin{figure}[th]
\vspace{1cm}
\centerline{\psfig{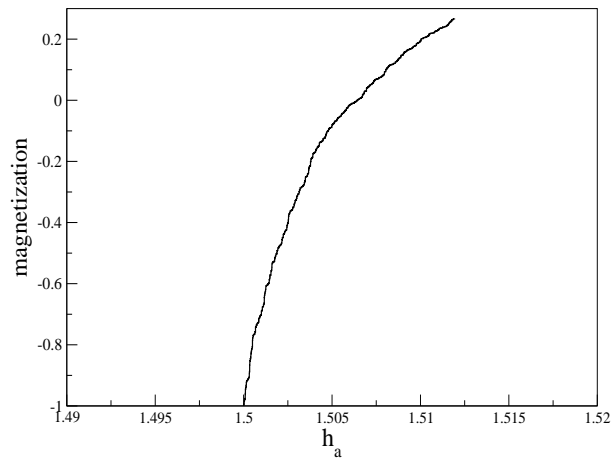}}
\vspace*{8pt}
\caption{The magnetization curve for $\Delta (=2.5) > \Delta_c $.
The magnetization increases in small irregular jumps as the field is 
increased.}
\end{figure}

\par For any  field, when the system has stopped evolving,
a typical distribution of the random fields over the unturned spins along the 
interface is as shown in
Fig.5. It consists of two steps. The higher step is for those sites 
that have large negative random fields that would require three upturned 
neighbors for its flipping. 
 The lower step corresponds to the random fields of those sites which
can flip if two of its neighbors have already flipped.
This shows that the the interface 
spins are strongly pinned.

\begin{figure}[t]
\vspace{1cm}
\begin{center}
\includegraphics[width=8cm,height = 6cm]{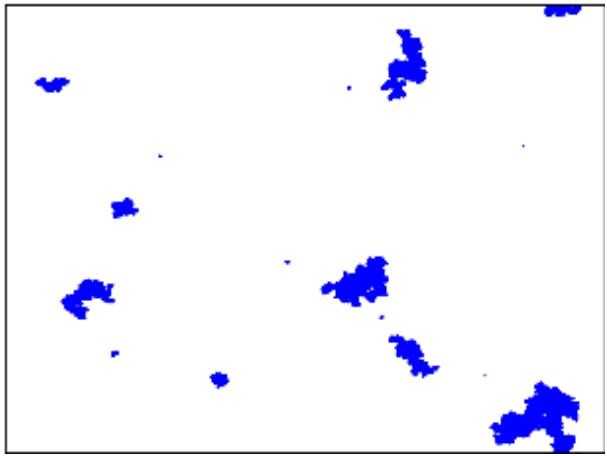}
\vspace*{8pt}
\caption {The domain of up spins (black) is shown for $\Delta (=2.5) > 
\Delta_c$, $h_a = h_{a}^{*} = 1.5000$.}
\end{center}
\end{figure}

\begin{figure}[th]
\vspace{1cm}
\centerline{\psfig{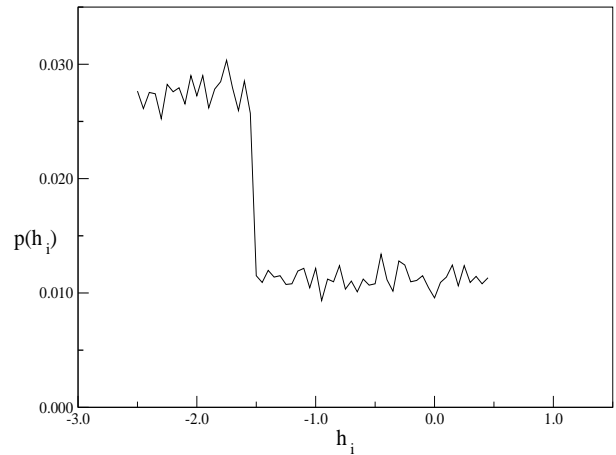}}
\vspace*{8pt}
\caption {The probability distribution of ``negative'' spins 
at a field when the growth of domains is arrested. This is for
$\Delta (= 2.5) > \Delta_c$.}
\end{figure}

The non-existence of
sites (at the interface) that can flip when one of its neighbor is up is 
responsible for the freezing of further spin flipping at the particular 
field.  

\begin{figure}[t]
\vspace{1cm}
\centerline{\psfig{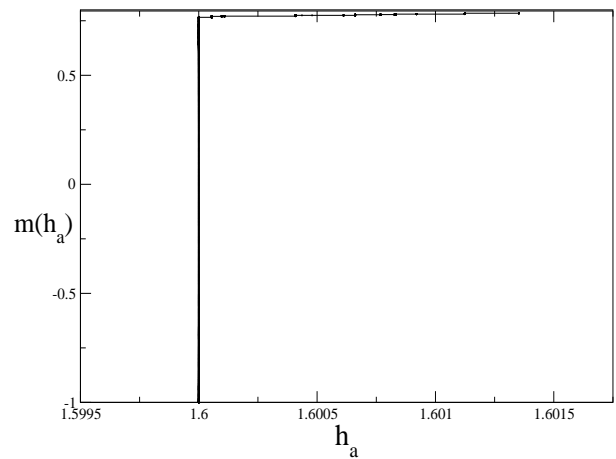}}
\vspace*{8pt}
\caption {The magnetization curve for $\Delta = \Delta_c = 2.4$.
The magnetization increases to about 0.76  in an infinite jump after which 
the field has to be increased for the magnetization to increase.}
\end{figure}

\item At the critical value of $\Delta = \Delta_c$ we see the growth of
a single domain 
and a discontinuous first order jump in the magnetization at
a particular value of the applied field. However, unlike 
the situation for $\Delta < \Delta_c$ here the domain of flipped spins 
encompasses bubbles of unturned spins
of various sizes at different stage  of its growth and the interface 
has overhangs and is fractal like. As a result magnetization
does not jump from -1 to +1 but to (around) 0.76
irrespective of the system size
(we have checked it for {\it L} = 500, 800, 2000,
4000, 5000 and 7000). The change in magnetization with field is shown in 
Fig.6.
\begin{figure}[th]
\vspace{1cm}
\centerline{\psfig{file=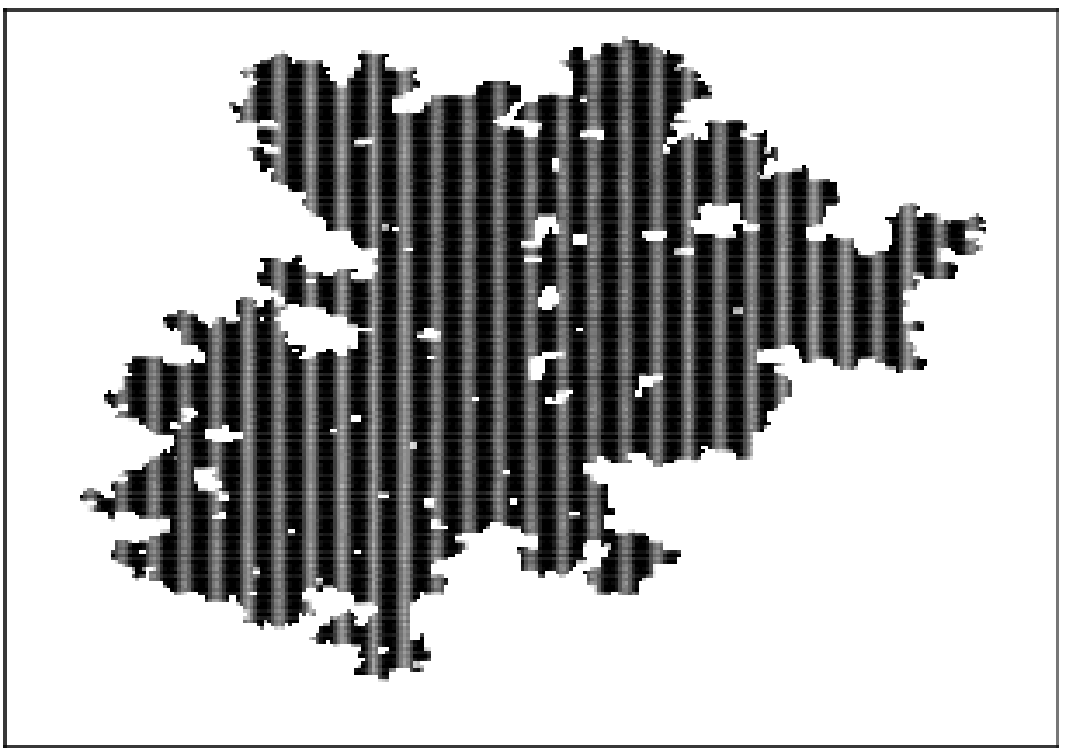,width = 8cm,height=6cm}}
\vspace*{8pt}
\caption {The domain of up spins (black) is shown for $\Delta = \Delta_c
(= 2.4)$, $h_a = h_{a}^{*} = 1.60000$.}
\end{figure}


\begin{figure}[t]
\vspace{1cm}
\begin{center}
\includegraphics[width=8cm,height = 6cm]{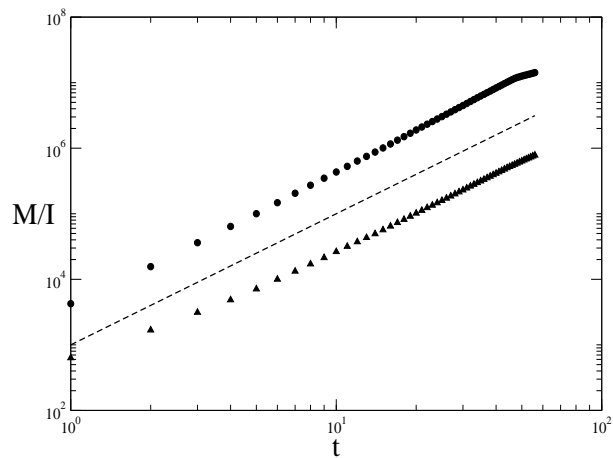}
\vspace{8pt}
\caption{The variation of the total number of spins,{\it M} and the number 
of spins along the interface I with time {\it t} (in units of 300 
Monte Carlo steps)for $\Delta = \Delta_c = 2.4$ and 
$h_a = h_{a}^{*} = 1.6000$.
Both M and I vary as the square of the 
time. The dotted line has a slope of unity and is a guide to the eye.}
\end{center}
\end{figure}
 A typical up-spin cluster at a certain time of its developement is 
shown in Fig.7. It shows the tortuous interface with bubbles of unturned 
spins (white region). In this case both the mass and interface develops as 
$M \sim I \sim t^2$ as is shown in Fig.8.

At any time during the growth of the up-spin domain the probability 
distribution of the random fields of the unturned spins along the   
interface has three steps as is shown in Fig.9. The lowest step corresponds
to the random fields on those sites  
that can flip if one of its neighbor 
is up. This guarantees the growth of domains as along the interface one 
neighbor will always be up.
However when the up-spins span the system the probability distribution 
of the random fields along the interface corresponds to those for the bubbles
only and is similar to that for $\Delta > \Delta_c$. This shows that the 
spins along the interface of the bubbles are strongly pinned. The external 
field has to be continuously increased for the magnetization to change from 
0.76 to +1 which will eventually fill up the bubbles \nolinebreak(Fig.6).

\begin{figure}[t]
\vspace{1cm}
\centerline{\psfig{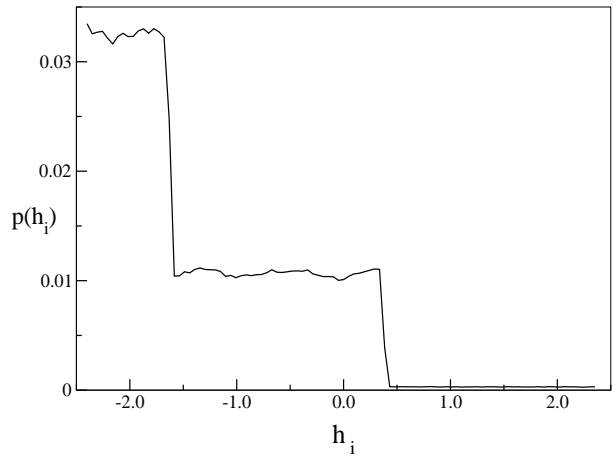}}
\vspace*{8pt}
\caption{Distribution  $p(h_i)$ of the random field $h_i$ over 
the spins along the interface between the up and the down spins for 
$\Delta = \Delta_c = 2.4$ and $h_a = 1.6000$. (The distribution 
remains invariant with time and is marked by three steps as is shown)}
\end{figure}

\end{romanlist}
 
\section{Conclusion}
We find a finite,non-zero $\Delta_c$ for bounded distribution.
This $\Delta_c$ demarcates a region ($\Delta > \Delta_c$) where infinite 
avalanches never happens to that from a region where magnetization shows a 
jump discontinuity from -1 to +1. Earlier studies indicate $\Delta_c = 0$ for 
unbounded distribution. 
We find at $\Delta_c$ the interface around the flipped domains have 
backbends and is 
fractal like. It encloses bubbles of unflipped spins in all length scales.

\end{document}